\documentclass[12pt]{article}

\usepackage{url}
\usepackage{bbm}
\usepackage{mathtools}
\usepackage{amssymb}
\usepackage{amsthm}
\usepackage{empheq}
\usepackage{latexsym}
\usepackage{enumitem}
\usepackage{eurosym}
\usepackage{dsfont}
\usepackage{appendix}
\usepackage{color} 
\usepackage[unicode]{hyperref}
\usepackage{frcursive}
\usepackage[utf8]{inputenc}
\usepackage[T1]{fontenc}
\usepackage{geometry}
\usepackage{multirow}
\usepackage{todonotes}
\usepackage{lmodern}
\usepackage{anyfontsize}
\usepackage{pgfplots}
\usepackage{stmaryrd}
\usepackage{float}
\usepackage{bookmark}
\usepackage{cleveref}
\usepackage{bm}

\graphicspath{ {images/} }

\pgfplotsset{compat=newest}

\definecolor{red}{rgb}{0.7,0.15,0.15}
\definecolor{green}{rgb}{0,0.5,0}
\definecolor{blue}{rgb}{0,0,0.7}
\hypersetup{colorlinks, linkcolor={red},citecolor={green}, urlcolor={blue}}
            
\makeatletter \@addtoreset{equation}{section}

\newtheorem{theorem}{Theorem}
\newtheorem{theorem2}{Theorem}[section]
\newtheorem{assumption}[theorem2]{Assumption}
\newtheorem{corollary}[theorem2]{Corollary}
\newtheorem{example}[theorem2]{Example}

\newtheorem{proposition}[theorem2]{Proposition}

\newtheorem{definition}[theorem2]{Definition}
\newtheorem{remark}[theorem2]{Remark}

\DeclareUnicodeCharacter{014D}{\=o}
\def \R{\mathbb{R}}

\newcommand{\norm}[1]{\left \lVert #1 \right \rVert_1}
\newcommand{\normsq}[1]{\left \lVert #1 \right \rVert_2}

\title{High-frequency dynamics of the implied volatility surface\footnote{This work benefits from the financial support of the Chaires Analytics and Models for Regulation, Deep Finance \& Statistics and Machine Learning \& systematic methods in finance. Bastien Baldacci gratefully acknowledges the financial support of the ERC Grant 679836 Staqamof. The author would like to than, Mehdi Tomas (\'Ecole Polytechnique), Mathieu Rosenbaum (\'Ecole Polytechnique) and Luxi Chen (Euronext) for numerous fruitful discussions, and especially, Iuliia Manziuk (\'Ecole Polytechnique) for her invaluable contributions.}}
\author{Bastien {\sc Baldacci}\footnote{\'Ecole Polytechnique, CMAP, 91128, Palaiseau, France,  bastien.baldacci@polytechnique.edu.} }

\begin{document}

\maketitle
\begin{abstract}
We present a Hawkes modeling of the volatility surface's high-frequency dynamics and show how the Hawkes kernel coefficients govern the surface's skew and convexity. We provide simple sufficient conditions on the coefficients to ensure no-arbitrage opportunities of the surface. Moreover, these conditions reduce the number of the kernel's parameters to estimate. Finally, we show that at the macroscopic level, the surface is driven by a sum of risk factors whose volatility processes are rough. 
\end{abstract}

\section{Introduction}

In 1973, the seminal paper of Black and Scholes \cite{black1973pricing} has drastically changed the options markets and their operation. In practice, even if their model is not always applied directly, being replaced by newer and more complex ones, the notions introduced thanks to this work are still very much in use. One of these concepts is, of course, the implied volatility, which is defined as the volatility such that the market price of an option coincides with its Black-Scholes counterpart. By finding a mapping of strikes and maturities to corresponding implied volatilities, one could build a volatility surface, which became one of the options market's crucial tools. \medskip

Empirical evidence from options markets shows that the surface exhibits a non-constant behavior, such as skew, smile, and a term structure, see, for example, \cite{derman1999regimes, derman1996local, engle2000testing}. In order to capture these features, the researchers and analysts started to develop more intricate models, mostly by changing the dynamics of the underlying asset: by introducing, for example, multiple factors and jumps to the diffusion. However, this way of modeling the volatility surface properties has two significant issues. First, the implied volatility surface can only be computed numerically in those frameworks, which is especially arduous for jump-diffusion models. Second, these models assume that the volatility surface dynamics are driven solely by the underlying; however, options markets have their supply-offer equilibrium, which cannot be described entirely by the underlying's dynamics, see, for example, \cite{bakshi2000call}. By considering the options' Delta as the only source of risk, these models fail to accurately reproduce the surface's dynamics, which leads to frequent recalibrations. Thus, it seems more relevant to model the dynamics of the implied volatility surface directly. 

\medskip
Another challenge in the volatility surface's modeling is the time scale. As we can see from the example of stocks' prices, the price process's behavior differs at thin and coarse time scales, and we want to take it into account for the volatility surface modeling. The diffusion models are usually used to obtain realistic dynamics for the underlying's volatility at coarse time scale. However, 
\begin{itemize}
\item The local volatility model reproduces the whole volatility surface at a given time but cannot reproduce the dynamics of the volatility observed in the market. 
\item Stochastic volatility models can accurately reproduce the volatility surface's dynamics but cannot fit both realized and implied volatility. In particular, the at-the-money skew of the volatility surface cannot be calibrated accurately using Brownian motion as a risk factor.
\item The rough volatility models, whose recent development has been motivated by the seemingly universal rough behavior of financial assets, capture key features of the implied volatility surface and its dynamics, see \cite{bayer2016pricing, el2019roughening, gatheral2018volatility}.\footnote{The log-volatility process of an asset is well-approximated by a fractional Brownian motion with small Hurst parameter $H\approx 0.1$.} However, as mentioned above, the calculation efforts to reproduce the volatility surface in this framework are considerably high.
\end{itemize}
The literature dealing with the implied volatility surface's direct modeling at the daily or more coarse time scale usually uses the principal component analysis to extract the volatility surface's main drivers, for example, \cite{cont2002dynamics, fengler2007semiparametric, kamal2010implied, skiadopoulos2000dynamics}. The resulting implied volatility surface is usually represented as a randomly fluctuating surface, whose deformation is driven by a small number of orthogonal random factors:
\vspace{-2mm}
\begin{itemize}
    \item The ``level'' factor corresponding to the global level of the whole surface of implied volatilities.
    \item The ``calendar'' factor corresponding to the skew of the volatility surface.
    \item The ``butterfly'' factor corresponding to the convexity of the volatility surface.
\end{itemize}
At the thin time scale, such as intraday, volatility surface dynamics models have been studied mainly from an empirical viewpoint, for example, in \cite{dunis2013forecasting, koopman2005forecasting, lee2019impacts, mayhew1995implied, wang2016information}. In these articles, the authors use time-series models on intraday high-frequency implied volatility data to forecast implied and realized volatility behavior. To the best of our knowledge, there is no model aimed at reproducing the intraday dynamics of the volatility surface, which is, even more, the case for thinner time scales as the high-frequency scale. \medskip

Because of the massive amount of high-frequency data, it is problematic to obtain an accurate estimation of volatility and covariance between two or more assets over a given period. Moreover, trading times and price processes are discrete, which makes diffusion models hardly usable. To solve these two issues,  we go on with a particular class of models, called tick-by-tick models, inspired by models of high-frequency asset prices using Hawkes processes. For example, in \cite{bacry2013modelling, bacry2014hawkes}, the authors consider a Hawkes processes-based model, where they assume that all price jumps are of the same size, the microscopic price of the asset is the difference of the number of upward and downward jumps, that is
\begin{align*}
    P_t = P_0 + N_t^+ - N_t^-, 
\end{align*}
where $(N^+,N^-)$ is a bi-dimensional Hawkes process with intensity kernel
\begin{align*}
    \boldsymbol{\phi}(t)  = 
    \begin{pmatrix}
     \phi^{++}(t) & \phi^{+-}(t) \\
     \phi^{-+}(t) & \phi^{--}(t)
    \end{pmatrix}.
\end{align*}
Set $\boldsymbol{\phi}$ is a set of endogenous sources of price moves: for example, $\phi^{+-}$ increases the intensity of upward price jumps after a downward price jump, creating a mean-reversion effect while $\phi^{++}$ creates a trending effect. This model reproduces accurately some stylized facts of the high-frequency price behavior, which can be incorporated into the Hawkes kernel: no-arbitrage property, bid-ask asymmetry, order flow long memory property, and a high degree of endogeneity of financial markets. Moreover, it enables to obtain closed-form expressions for important statistical measures of high-frequency financial data, such as the signature plot or Epps effect. This model has been extended to the multi-asset case in \cite{bacry2013some}, where the authors show the reproduction of the lead-lag effect between two assets, more general multi-asset case is studied in \cite{tomas2019microscopic}. Finally, note that one important advantage of the Hawkes-based models is that their long-term behavior is consistent with financial assets' roughness. At coarse time scale, the price process defined above converges to a stochastic rough volatility process. This property has raised interest in building microscopic models for market dynamics that reproduce rough volatility at a macroscopic scale, see \cite{jaisson2015limit, jaisson2016rough, jusselin2018no, tomas2019microscopic}. \medskip

In this paper, we propose a tick-by-tick model for the high-frequency dynamics of the volatility surface. Assuming $\#\mathcal{K}$ strikes and $\#\mathcal{T}$ maturities, the microscopic volatility surface is modeled as a $\#\mathcal{K}\times \#\mathcal{T}$-dimensional process $(\boldsymbol{\sigma}_t^{(k,\tau)})_{(k,\tau)\in \mathcal{K}\times\mathcal{T}}$, where 
\begin{align*}
    \sigma_t^{k,\tau} = N_t^{(k,\tau)+}-N_t^{(k,\tau)-},
\end{align*}
and $(N^{(k,\tau)+},N^{(k,\tau)-})_{(k,\tau)\in\mathcal{K}\times\mathcal{T}}$ is a $2\times\#\mathcal{K}\times \#\mathcal{T}$-dimensional Hawkes process. The processes $N_t^{(k,\tau)+}$ (resp. $N_t^{(k,\tau)-}$) count the number of upward (resp. downward) moves of the implied volatility of option with strike $k$ and maturity $\tau$. This modelling is especially well suited to Foreign Exchange (FX) options markets, where options are directly quoted in terms of implied volatility.\footnote{It can also be used on markets where options are quoted in prices, by assuming for example sticky moneyness of the volatility surface.} \medskip

We show how the coefficients of the Hawkes kernel govern the skew and convexity of the surface. Moreover, they can be parametrized in a way that no-arbitrage conditions of the volatility are satisfied. In addition, this parametrization of the volatility surface reduces the number of parameters to calibrate. We also study classic Hawkes kernels such as power-law, and more precisely, the kernel's parameters influence on the shape of the surface. The conditions provide practical ways to parametrize the volatility surface, reproducing most of the market's stylized facts. \medskip

To verify the model's properties at large time scales, we use the limit theorems as in \cite{tomas2019microscopic} to show, for example, that our volatility surface behaves like a diffusion process. We obtain that a sum of orthogonal factors drives the volatility surface dynamics, whose volatility processes are rough. For example, the highest eigenvalue of $\boldsymbol{\phi}$ is the ``level'' of the volatility surface. Hence, our results at the macroscopic scale are in the spirit of \cite{cont2002dynamics}, except that the risk factors driving the volatility surface come from the microscopic interactions between market participants. We show that our microscopic model can be used as a backtesting environment for trading and market-making strategies on options. At the macroscopic time scale, it can be used to compute these strategies' impact on the volatility surface. 

\medskip
The article is structured as follows. In Section \ref{sec_micro_model}, we introduce the model for the high-frequency dynamics of the implied volatility surface and provide sufficient conditions to ensure absence of arbitrage. In Section \ref{sec_macro_volatility}, we study the macroscopic behavior of the volatility surface. Finally, Section \ref{sec_applications} is devoted to the numerical applications of the model to option market-making strategies.

\section{Microscopic modelling of the volatility surface}\label{sec_micro_model}

\subsection{Framework and no-arbitrage conditions}

In this section, we first formalize our tick-by-tick model for the microscopic dynamics of the volatility surface. Let us consider a time horizon $T>0$ and a set of options on an underlying asset with strikes $\{k_1, \dots, k_N\}$ and maturities $\{\tau_1,\dots,\tau_M\}$, $N,M\in \mathbb{N}^\star$.\footnote{The strikes are expressed in log-moneyness, that is $k_i=\log(\frac{K_i}{S})$ where $S$ is the price of the underlying asset and $K_i>0$ for $i\in \{1,\dots,N\}$.} At the microscopic scale, the point $(k,\tau)$ of the volatility surface is a piecewise constant process with upward and downward jumps defined by the following tick-by-tick model: 
\begin{align*}
    \sigma_t^{(k,\tau)} = N_t^{(k,\tau)+} - N_t^{(k,\tau)-}, \quad t\geq 0, (k,\tau)\in \{k_1,\dots,k_N\}\times \{\tau_1,\dots,\tau_M\} = \mathcal{K}\times\mathcal{T}, 
\end{align*}
where $N_t^{(k,\tau)+}$ counts the number of upward volatility moves,
and $N_t^{(k,\tau)-}$ counts the number of downward volatility moves. The vector $\mathbf{N} = (N^{(k,\tau)+}, N^{(k,\tau)-})_{(k,\tau)\in \mathcal{K}\times\mathcal{T}}$ is a $2\times\#\mathcal{K}\times\#\mathcal{T}$-dimensional Hawkes process with intensity 
\begin{align*}
    \boldsymbol{\lambda}_t = \boldsymbol{\mu}_t + \int_0^t \boldsymbol{\phi}(t-s)d\mathbf{N}_{s}.
\end{align*}
\begin{definition}
The microscopic volatility surface is defined by the $\mathcal{M}_{\#\mathcal{K}\times\#\mathcal{T}}(\mathbb{R})-$valued process $\boldsymbol{\sigma_t} = \big(\sigma_t^{(k,\tau)}\big)_{t\geq 0,(k,\tau)\in \mathcal{K}\times\mathcal{T}}$.
\end{definition}
This definition is related to the notion of market impact: a net buy pressure on option $(k,\tau)$ leads to a higher implied volatility $\sigma^{(k,\tau)}$ and conversely for a net sell pressure. This tick-by-tick modelling is commonly used for price processes, see for example \cite{el2018microstructural, tomas2019microscopic}. As stated in the introduction, the use of a tick-by-tick model for the moves of the volatility surface is particularly well suited to FX options markets, where the options are directly quoted in terms of implied volatility. For other markets, as the price of an option is an increasing bijection of the implied volatility function, the impact on prices and implied volatility is qualitatively the same. 
\begin{remark}
Throughout the paper, we assume implicitly that the volatility process of an option $(k,\tau)$ does not make unitary jumps but rather jumps equal to a fixed tick size. This tick size is assumed to be the same for all strikes and maturities for ease of notation. However, this assumption can be relaxed without a change of the key results. 
\end{remark}

Let the vector $\boldsymbol{\mu}: \mathbb{R}_+ \to \mathbb{R}_+^{\#\mathcal{K}\times\#\mathcal{T}}$ denote the baseline (that is the intensity vector without self and mutual excitation) and $\boldsymbol{\phi}: \mathbb{R}_+ \to \mathcal{M}_{2\times\#\mathcal{K}\times\#\mathcal{T}}(\mathbb{R})$ denote the kernel encoding the self- and cross-excitation of the different points of the volatility surface. These terms deserve some financial interpretations:
\begin{itemize}
    \item The process $\mu^{(k,\tau)+}(t)$ (resp. $\mu^{(k,\tau)-}(t)$) is the exogenous source of upward (resp. downward) moves of the volatility surface at point $(k,\tau)$.  For example, the exogenous sources of jumps corresponding to at-the-money options have higher values than far-from-the-money options. 
    \item The kernel $\boldsymbol{\phi}$ describes the endogenous influence of the past volatility moves to the current intensities, namely self- and cross-excitement. We denote by $\phi^{(k,\tau)s,(\tilde k,\tilde \tau)\tilde{s}}$ the influence of $N^{(\tilde k,\tilde \tau)\tilde{s}}$ on $N^{(k,\tau)s}$, $(s, \tilde s)\in \{+,-\}^2$. For example, the quantity $\phi^{(k_i,\tau_j)+,(k_i,\tau_j),+}$ has a trending self-exciting role: it increases the intensity of upward volatility jumps of option $(k_i,\tau_j)$ after an upward volatility jump of the same option $(k_i,\tau_j)$, similarly for $\phi^{(k_i,\tau_j)-,(k_i,\tau_j),-}$. Or, for instance, for $(i_1,i_2),(j_1,j_2)$, the quantity  $\phi^{(k_{i_1},\tau_{j_1})+,(k_{i_2},\tau_{j_2}),-}$ has a mean-reverting cross-exciting role: it increases the intensity of upward volatility jumps of option $(k_{i_1},\tau_{j_1})$ after a downward volatility jump of the option $(k_{i_2},\tau_{j_2})$. 
\end{itemize}
The kernel $\boldsymbol{\phi}$ models the shape of the volatility surface at the microscopic scale. The following two examples aim at describing how the shape of the volatility surface is linked to the Hawkes kernel. 
\begin{example}\label{example_skew}
\textbf{Volatility skew.} Let us take a single slice of volatility,\footnote{In this article, we refer to  the set of options having the same maturity as the slice of volatility.} that is $\mathcal{T}=\{\tau\}$, and, for the sake of clarity, consider only two options $(k_1,\tau),(k_2,\tau)$. Let us assume a volatility skew, typical for equity markets, that is for all $t\geq0$, \[\mathbb{E}[\sigma_t^{(k_1,\tau)}] \geq \mathbb{E}[\sigma_t^{(k_2,\tau)}]\] for $k_1\leq k_2$. Moreover, let us assume the following kernel structure for the $4$-dimensional Hawkes process:
\begin{align*}
\boldsymbol{\phi} = 
\begin{pmatrix}
    0 & 0 & \phi^{(k_1,\tau)+,(k_2,\tau)+} & \phi^{(k_1,\tau)+,(k_2,\tau)-} \\
    0 & 0 & \phi^{(k_1,\tau)-,(k_2,\tau)+} & \phi^{(k_1,\tau)-,(k_2,\tau)-} \\
    \phi^{(k_2,\tau)+,(k_1,\tau)+} & \phi^{(k_2,\tau)+,(k_1,\tau)-} & 0 & 0 \\
    \phi^{(k_2,\tau)-,(k_1,\tau)+} & \phi^{(k_2,\tau)-,(k_1,\tau)-} & 0 & 0
\end{pmatrix},
\end{align*}
where we drop the time dependence of the kernel coefficients for the sake of readability. In this example we assume no self- or cross-excitation of the volatility processes on the same option, that is
\[\phi^{(k_i,\tau)s_1,(k_i,\tau)s_2} =0, \quad i\in \{1,2\}, s_i\in\{+,-\},\]
but only cross-excitation from option $(k_1,\tau)$ to option $(k_2,\tau)$ and conversely. Assume a high number of buy orders on option $(k_2,\tau)$. This increase of the counting process $N^{(k_2,\tau)+}$ increases the level of volatility $\sigma^{(k_2,\tau)}$. However, it also impacts the left side of the volatility skew because of the coefficients $\phi^{(k_1,\tau)-,(k_2,\tau)+},\phi^{(k_1,\tau)+,(k_2,\tau)+}$. The slice of volatility can be impacted in the following ways:
\begin{itemize}
    \item If $\phi^{(k_1,\tau)-,(k_2,\tau)+}>\phi^{(k_1,\tau)+,(k_2,\tau)+}$, then, on average, an increase of the volatility for option $(k_2,\tau)$ decreases the volatility for option $(k_1,\tau)$ so that the volatility slice for maturity $\tau$ flattens. 
    \item If $\phi^{(k_1,\tau)-,(k_2,\tau)+}<\phi^{(k_1,\tau)+,(k_2,\tau)+}$, then, on average, an increase of the volatility for option $(k_2,\tau)$ increases the volatility for option $(k_1,\tau)$:
    \begin{itemize}
        \item If the buy orders on option $(k_2,\tau)$ lead to more buy orders on option $(k_1,\tau)$, then the volatility skew steepens.
        \item If the buy orders on option $(k_2,\tau)$ lead to less buy orders on option $(k_1,\tau)$, then the volatility skew flattens. 
    \end{itemize} 
\end{itemize}
These cases are represented in Figure \ref{fig::skew} below. 
\begin{figure}[H]
\centering
\includegraphics[width=17cm]{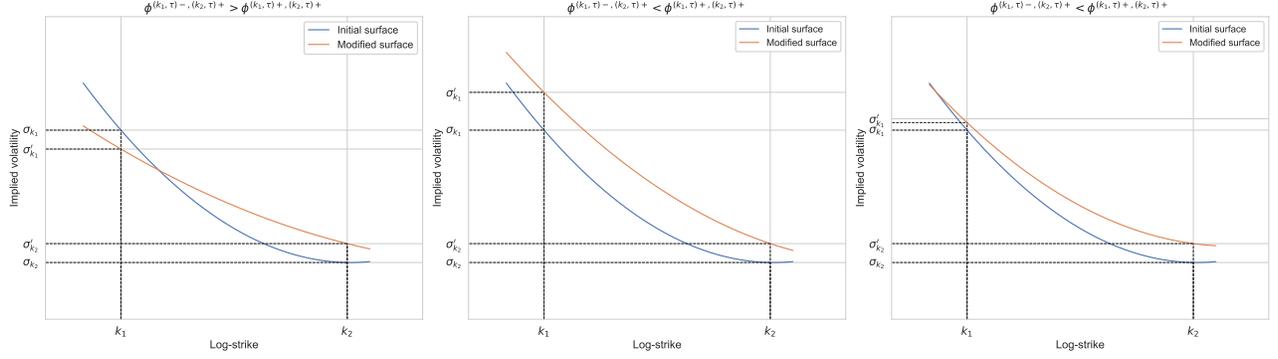}
\caption{Evolution of the slice of volatility when $\phi^{(k_1,\tau)-,(k_2,\tau)+}>\phi^{(k_1,\tau)+,(k_2,\tau)+}$ (left), $\phi^{(k_1,\tau)-,(k_2,\tau)+}<\phi^{(k_1,\tau)+,(k_2,\tau)+}$ and there is less buy orders on option $(k_2,\tau)$ (center), $\phi^{(k_1,\tau)-,(k_2,\tau)+}<\phi^{(k_1,\tau)+,(k_2,\tau)+}$ and there is more buy orders on option $(k_2,\tau)$ (right)}
\label{fig::skew}
\end{figure}
\end{example}

\begin{example}
\textbf{Volatility convexity.} Let us take a slice of volatility, that is $\mathcal{T}=\{\tau\}$, and consider only three options $(k_1,\tau),(k_2,\tau),(k_3,\tau)$. Assume a volatility smile, as in FX markets, for example, such that for all $t\geq0$, $$\mathbb{E}[\sigma_t^{(k_1,\tau)}] \geq \mathbb{E}[\sigma_t^{(k_2,\tau)}], \quad \mathbb{E}[\sigma_t^{(k_2,\tau)}] \leq \mathbb{E}[\sigma_t^{(k_3,\tau)}]$$ for $k_1\leq k_2 \leq k_3$. Thus, we consider that $k_1$ is an in-the-money strike, $k_2$ is an at-the-money strike and $k_3$ is an out-of-the-money strike. Moreover, assume the following kernel structure for the $6$-dimensional Hawkes process:
\begin{align*}
\mathbf{\phi} = 
\begin{pmatrix}
    0 & 0 & 0 & 0 & \phi^{(k_1,\tau)+,(k_3,\tau)+} & \phi^{(k_1,\tau)+,(k_3,\tau)-} \\
    0 & 0 & 0&  0 & \phi^{(k_1,\tau)-,(k_3,\tau)+} & \phi^{(k_1,\tau)-,(k_3,\tau)-} \\
    0 & 0 & 0 & 0 & 0 & 0 \\
    0 & 0 & 0 & 0 & 0 & 0 \\
    \phi^{(k_3,\tau)+,(k_1,\tau)+} & \phi^{(k_3,\tau)+,(k_1,\tau)-} & 0 & 0 & 0 & 0\\
    \phi^{(k_3,\tau)-,(k_1,\tau)+} & \phi^{(k_3,\tau)-,(k_1,\tau)-} & 0 & 0 & 0 & 0
\end{pmatrix}.
\end{align*}
In this example we assume no self- and cross-excitation of the volatility processes on the same options, and no self or cross-excitation coming from the at-the-money option $(k_2,\tau)$. The only cross-excitation comes from the out-of-the-money to the in-the-money option and conversely.  The at-the-money part of the smile is not-affected by far-from-the-money transactions for sake of clarity of this example. We assume a high number of sell orders on $(k_3,\tau)$. 
\begin{itemize}
    \item If $\phi^{(k_1,\tau)-,(k_3,\tau)-}>\phi^{(k_1,\tau)+,(k_3,\tau)-}$, the convexity of this slice of volatility decreases on average. 
    \item If $\phi^{(k_1,\tau)-,(k_3,\tau)-}<\phi^{(k_1,\tau)+,(k_3,\tau)-}$, the volatility smile becomes (on average) left skewed. 
\end{itemize}
Finally, assume a high number of buy orders on the out-of-the-money option $(k_3,\tau)$. This increase of the counting process $N^{(k_3,\tau)+}$ increases the level of volatility $\sigma^{(k_3,\tau)}$, and it also impacts the left side of the volatility smile. 
\begin{itemize}
    \item If $\phi^{(k_1,\tau)-,(k_3,\tau)+}<\phi^{(k_1,\tau)+,(k_3,\tau)+}$, left and right side of the volatility smile rise (on average) so that the convexity of this slice of volatility increases. 
    \item If $\phi^{(k_1,\tau)-,(k_3,\tau)+}>\phi^{(k_1,\tau)+,(k_3,\tau)+}$, the volatility smile becomes right-skewed.
\end{itemize}
These four cases are represented in Figure \ref{fig::smile} below. 
\begin{figure}[H]
\centering
\includegraphics[width=17cm]{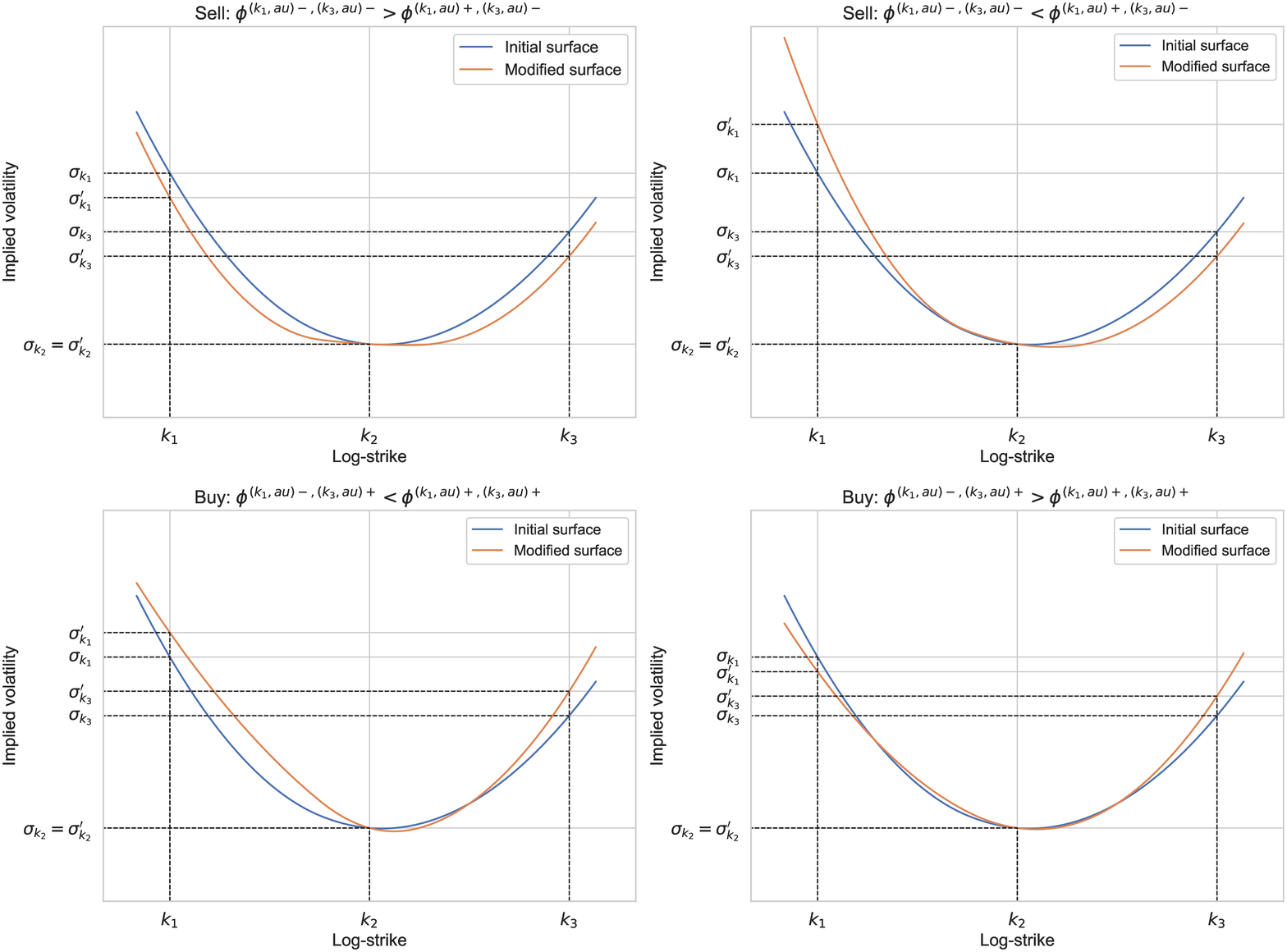}
\caption{Evolution of the slice of volatility with sell orders on $(k_3,\tau)$ and $\phi^{(k_1,\tau)-,(k_3,\tau)-}>\phi^{(k_1,\tau)+,(k_3,\tau)-}$ (upper left), $\phi^{(k_1,\tau)-,(k_3,\tau)-}<\phi^{(k_1,\tau)+,(k_3,\tau)-}$ (upper right). Evolution of the slice of volatility with buy orders on $(k_3,\tau)$ and $\phi^{(k_1,\tau)-,(k_3,\tau)+}<\phi^{(k_1,\tau)+,(k_3,\tau)+}$ (lower left), $\phi^{(k_1,\tau)-,(k_3,\tau)+}>\phi^{(k_1,\tau)+,(k_3,\tau)+}$ (lower right). }
\label{fig::smile}
\end{figure}
\end{example}
These two simple examples show that a suitable parametrization for a multidimensional Hawkes process can reproduce shapes of a tick-by-tick volatility surface. However, to ensure meaningful results, it is crucial to enforce no-arbitrage conditions on the volatility surface at the microscopic scale. Following \cite{gatheral2014arbitrage}, we consider that a volatility surface is arbitrage-free if it is free of calendar spread and butterfly arbitrages. We now explain these two notions and show how to incorporate these properties into our framework. \medskip

A calendar spread arbitrage is an implied volatility arbitrage consisting in buying and selling two calls (or puts) on the same underlying with the same strike and different maturities. As the time value of the option with a shorter maturity decreases quicker than the one with a longer maturity, the latter should have a higher price than the former. Thus, there is no arbitrage opportunity in a calendar spread strategy if European call option prices are monotonous with respect to the maturity. Assuming that dividends are proportional to the price of the underlying, a necessary and sufficient condition for the volatility surface to be free of calendar spread arbitrage is the following.
\begin{proposition}\label{def_calendar_spread}
Let us define the total implied variance at time $t$ for an option $(k,\tau)$ as $\varpi_t(k,\tau)= (\sigma^{(k,\tau)}_t)^2 \tau$. The microscopic volatility surface is
free of calendar spread arbitrage if and only if
\begin{align*}
    \partial_\tau \varpi_t(k,\tau)\geq 0 \text{ for all } (t,k,\tau)\in[0,T]\times\mathcal{K}\times\mathcal{T}.
\end{align*}
\end{proposition}
The proof can be found in \cite{gatheral2014arbitrage}. This condition cannot be applied directly to our model as we work with a discrete implied volatility. To incorporate the no calendar spread arbitrage condition in the Hawkes kernel, we rewrite the condition of Proposition \ref{def_calendar_spread} as
\begin{align*}
 \dfrac{(\sigma^{(k,\tau_i)}_t)^2 \tau_i - (\sigma^{(k,\tau_j)}_t)^2 \tau_j}{\tau_i - \tau_j} \geq 0, \quad \text{ for all } k\in\mathcal{K}, (\tau_i,\tau_j)\in \mathcal{T}^2.
\end{align*}
We recall that, for a given Hawkes process $N^{(k,\tau)s}$ with intensity $\lambda^{(k,\tau)s}$, $s \in \{+, -\}$, we have the following equalities:
\begin{align}\label{relationship_hawkes}
\begin{split}
    & \mathbb{E}\big[N_t^{(k,\tau)s}\big] = \int_0^t \mathbb{E}\big[\lambda^{k,\tau,s}_u\big] du, \\
    & \mathbb{E}\big[\lambda^{(k,\tau)s}_t\big] = \mu^{(k,\tau)s} + \sum_{(\tilde k,\tilde \tau)}\sum_{\tilde{s}\in \{+,-\}}\int_0^t \phi^{(k,\tau)s,(\tilde k,\tilde \tau)\tilde{s}}(t-u)\mathbb{E}\big[\lambda^{(\tilde{k},\tilde \tau)\tilde s}_u\big]du,   
\end{split}
\end{align}
where, for sake of simplicity, we assume that the exogenous source of jumps $\boldsymbol\mu$ is constant. Assuming $\tau_i \geq \tau_j$ and using the definition of the expectation of a Hawkes process, for no statistical arbitrage for calendar spread we obtain
\begin{align}\label{ineq_calendar_spread}
 \hspace{-3mm}   \mathbb{E}\big[\mu^{(k,\tau_i)+} \!\! - \! \mu^{(k,\tau_i)-}\big] \!\!+\!\! \int_0^t\! \mathbb{E}[\lambda^{(k,\tau_i)+}_s\!\! -\! \lambda^{(k,\tau_i)-}_s\!] ds \!\geq\!\! \sqrt{\frac{\tau_j}{\tau_i}}\!\Big(\mathbb{E}\big[\mu^{(k,\tau_j)+}\!\!-\!\mu^{(k,\tau_j)-}\big]\!\! +\!\! \int_0^t\! \mathbb{E}[\lambda^{(k,\tau_j)+}_s\!\!-\!\lambda^{(k,\tau_j)-}_s\!] ds \Big). 
\end{align}
Using Equation \eqref{relationship_hawkes}, we can deduce that to satisfy the above condition we can directly impose the following constraints on each coefficient of the intensities:
\begin{align}\label{calendar_spread_sufficient_cond}
\begin{split}
     & \boldsymbol{\phi^{(k,\tau_i)+, \cdot}} =\beta^{\phi,+} \sqrt{\dfrac{\tau_j}{\tau_i}} \boldsymbol{\phi^{(k,\tau_j)+, \cdot}}, \quad 
    \boldsymbol{\phi^{(k,\tau_i)-, \cdot}} = \beta^{\phi,+}\sqrt{\dfrac{\tau_j}{\tau_i}} \boldsymbol{\phi^{(k,\tau_j)-, \cdot}}, \\
    & \mu^{(k,\tau_i)+} = \beta^{\mu,+}\sqrt{\dfrac{\tau_j}{\tau_i}}\mu^{(k,\tau_j)+},\quad  \mu^{(k,\tau_i)-} = \beta^{\mu,-}\sqrt{\dfrac{\tau_j}{\tau_i}}\mu^{(k,\tau_j)-},
\end{split}
\end{align}
where $(\beta^{\mu,+},\beta^{\mu,-},\beta^{\phi,+},\beta^{\phi,-})\in (1,+\infty)^4$. This leads to the following corollary.
\begin{corollary}
The microscopic volatility surface $(\boldsymbol{\sigma_t})_{t\geq0}$ is free of statistical calendar spread arbitrage if condition \eqref{calendar_spread_sufficient_cond} is satisfied. 
\end{corollary}
We emphasize that Condition \eqref{calendar_spread_sufficient_cond} is a sufficient but not necessary condition to ensure absence of statistical calendar spread arbitrage. Indeed, different kernel parametrizations are possible, notably by imposing the conditions on $\boldsymbol{\psi} = \sum_{l \geq 1} \boldsymbol{\phi}^{*l}$, where $\boldsymbol{\phi}^{*l}$ is the $l$-th convolution product of $\boldsymbol{\phi}$ with itself. Moreover, the coefficients $(\beta^{\mu,+}, \beta^{\mu,-}, \beta^{\phi,+}, \beta^{\phi,-})$ can be time- or strike-dependent for each coefficient of the Hawkes kernel. \medskip

Condition \eqref{calendar_spread_sufficient_cond} leads to a maturity parametrization for the Hawkes kernel, thus reduces the number of parameters to calibrate. For example, let us assume $\beta^{\mu,+}=\beta^{\mu,-}=\beta^{\phi,+}=\beta^{\phi,-}=1$, so that the no calendar spread arbitrage constraint \eqref{ineq_calendar_spread} becomes an equality. This way, we have only $2\times \#\mathcal{K}$ coefficients of the Hawkes kernel to calibrate instead of $2\times \#\mathcal{K}\times\#\mathcal{T}$. \medskip

Another condition for the volatility surface to be arbitrage-free is the absence of butterfly arbitrage. In financial markets, a butterfly is a strategy that consists in: buying one call option with strike $k_1$ and one call option with strike $k_3$, and selling two call options with strike $k_2$ with the same maturity $\tau$, and $k_1<k_2<k_3, k_2-k_1=k_3-k_2$. If we denote by $C(k, \tau)$ the price of a call option with strike $k$ and maturity $\tau$, the payoff of the butterfly strategy is 
\begin{align*}
    BF_{(k_1,k_2,k_3),\tau} =C(k_1,\tau) - 2 C(k_2,\tau) + C(k_3,\tau). 
\end{align*}
Using the results of \cite{breeden1978prices}, we know that if the price of a vanilla option is available, one can retrieve the density of the underlying $f_S$ via the formula
\begin{align*}
  f_S(k,\tau) = \partial_{kk} C(k,\tau), \text{for all }(k,\tau)\in \mathcal{K}\times\mathcal{T}.
\end{align*}
Without conditions on call option prices (and therefore on the implied volatility), $f_S$ in practice is not necessarily a density. The butterfly strategy is related to the function $f_S$ in the following way: for a set of strikes $(k_1,\dots,k_N)$ and a maturity $\tau$, $f_S(k_i,\tau)\approx \frac{C(k_{i-1},\tau) -2C(k_i,\tau)+C(k_{i+1},\tau)}{(k_i-k_{i-1})^2}$ for $i\in \{1,\dots,N-1\}$. Therefore, the absence of butterfly arbitrage states that $f_S$ must stay positive and integrate to one. The following proposition, whose proof can be found in \cite{gatheral2014arbitrage}, provides necessary and sufficient conditions for the volatility surface to be free of butterfly arbitrage.
\begin{proposition}
Let us define for a slice $\tau$ the following quantities at time $t\in[0,T]$
\begin{align*}
   & d_t(k,\tau) = -\frac{k}{\sqrt{\varpi_t(k,\tau)}}+ \frac{\sqrt{\varpi_t(k,\tau)}}{2} , \\
   & g_t(k,\tau)= \Big(1- \frac{\partial_k\varpi_t(k,\tau)}{2\varpi_t(k,\tau)}\Big)^2 - \frac{\partial_k\varpi_t(k,\tau)^2}{4}\Big(\frac{1}{\varpi_t(k,\tau)}+\frac{1}{4}\Big)+ \frac{\partial_{kk}\varpi_t(k,\tau)}{2}. 
\end{align*}
A slice is free of butterfly arbitrage if and only if $g_t(k,\tau)\geq 0$ and $\lim_{k\to +\infty}d_t(k,\tau)=-\infty$. 
\end{proposition}
This condition is more complicated to express in terms of the Hawkes kernel. The limit over $d(k,\tau)$ when $k\to +\infty$ implies that the density of the underlying integrates to one, and can be rewritten as
\begin{align*}
    (\sigma^{(k,\tau)})^2 < \frac{2k}{\tau},
\end{align*}
for large $k$, which is highly similar to the Roger-Lee moment formula for tail behavior of the volatility surface, see \cite{lee2004moment}. For technical convenience, we can impose the stricter condition
\begin{align*}
    \sigma^{(k_i,\tau)} = \beta \sqrt{\frac{k_i}{ k_j}}\sigma^{(k_j,\tau)},\quad \beta\in (0,1),
\end{align*}
and using the results of \eqref{relationship_hawkes}, we can deduce that this relationship is satisfied by imposing the following condition on the coefficients of the Hawkes kernel:
\begin{align}\label{cond_no_arbitrage_butterfly_kernel}
\begin{split}
     & \boldsymbol{\phi^{(k_i,\tau)+, \cdot}} =\beta \sqrt{\dfrac{k_i}{k_j}} \boldsymbol{\phi^{(k_j,\tau)+, \cdot}}, \quad 
    \boldsymbol{\phi^{(k_i,\tau)-, \cdot}} = \beta\sqrt{\dfrac{k_i}{k_j}} \boldsymbol{\phi^{(k_j,\tau)-, \cdot}}, \\
    & \mu^{(k_i,\tau)+} = \beta\sqrt{\dfrac{k_i}{k_j}}\mu^{(k_j,\tau)+},\quad  \mu^{(k_i,\tau)-} = \beta\sqrt{\dfrac{k_i}{k_j}}\mu^{(k_j,\tau)-},
\end{split}
\end{align}
for large $(k_i,k_j)\in \mathcal{K}^2,\tau\in\mathcal{T}$. In practice, $\beta\in(0,1)$ must be linked to the number of finite moments of the underlying asset, see \cite{lee2004moment}. The condition $g_t(k,\tau)\geq 0$, which guarantees the positivity of the density of the underlying asset, is harder to include. We propose in the following section an alternative to this condition which ensures at-the-money convexity of the volatility slices. This alternative is also more suited in our case of discrete volatility surfaces.

\subsection{Parametrization of the volatility surface}

\subsubsection{Three-points volatility surface}

An important advantage of this kind of microscopic volatility modeling is that it is easy to control right and left skew and convexity of the slices of volatility. We choose one slice $\tau$ and we note that, by using no calendar spread arbitrage condition \eqref{calendar_spread_sufficient_cond}, it is sufficient to parametrize only one slice of volatility. We take three strikes of the options $-25\Delta P, 50\Delta$ and $25\Delta C$, that is
\begin{itemize}
    \item the put option $-25\Delta P$ has a strike such that its delta is equal to $-0.25$,
    \item the call option $50\Delta $ has a strike such that its delta is equal to $0.5$,
    \item the call option $25\Delta C$ has a strike such that its delta  is equal to $0.25$.
\end{itemize}
We denote by $\sigma^{-25\Delta P,\tau}, \sigma^{50\Delta, \tau}, \sigma^{25\Delta C, \tau}$ the volatilities associated to these options. The implied volatilities of the so-called risk-reversal and butterfly strategies, which are in fact measures of the skew and the convexity of the slice, with maturity $\tau$ are given by
\begin{align*}
    BF_{25\Delta,\tau} = \frac{\sigma^{25\Delta C,\tau}+\sigma^{-25\Delta P,\tau}}{2}-\sigma^{50\Delta,\tau}, \quad RR_{25\Delta,\tau} = \sigma^{25\Delta C,\tau}- \sigma^{-25\Delta P,\tau}. 
\end{align*}
In particular, the convexity of the volatility slice is verified if and only if
\begin{align*}
    \frac{\mathbb{E}[\sigma_t^{25\Delta C,\tau}]+\mathbb{E}[\sigma_t^{-25\Delta P,\tau}]}{2}\geq \mathbb{E}[\sigma_t^{50\Delta,\tau}]. 
\end{align*}
If this condition is violated, one can find an arbitrage opportunity by creating a butterfly product whose price is negative. It can be replaced by a more restrictive condition:
\begin{align*}
    \frac{\mathbb{E}[\sigma_t^{25\Delta C,\tau}]+\mathbb{E}[\sigma_t^{-25\Delta P,\tau}]}{2} =  \beta^B\mathbb{E}[\sigma_t^{50\Delta,\tau}], \quad \beta^B \in [1,+\infty).
\end{align*}
By imposing conditions directly on the coefficients of the Hawkes kernel, we obtain 
\begin{align*}
     \frac{1}{2}\sum_{(\tilde k,\tilde \tau)}\sum_{s\in \{+,-\}}& \big(\phi^{(25\Delta C,\tau)+,(\tilde k,\tilde \tau)s} + \phi^{(-25\Delta P,\tau)+,(\tilde k,\tilde \tau)s}-\phi^{(25\Delta C,\tau)-,(\tilde k,\tilde \tau)s} - \phi^{(-25\Delta P,\tau)-,(\tilde k,\tilde \tau)s}\big) \\
     & = \beta^B \sum_{(\tilde k,\tilde \tau)}\sum_{s\in \{+,-\}} \phi^{(50\Delta ,\tau)+,(\tilde k,\tilde \tau)s}, \\
\end{align*}
which leads to the following natural condition to ensure convexity:
\begin{align}\label{condition_butterfly_convexity_atm}
\begin{split}
    &\frac{\phi^{(25\Delta C,\tau)+,(\tilde k,\tilde \tau)s} + \phi^{(-25\Delta P,\tau)+,(\tilde k,\tilde \tau)s}}{2} = \beta^B \phi^{(50\Delta ,\tau)+,(\tilde k,\tilde \tau)s}, \\
    & \frac{\phi^{(25\Delta C,\tau)-,(\tilde k,\tilde \tau)s} + \phi^{(-25\Delta P,\tau)-,(\tilde k,\tilde \tau)s}}{2} = \beta^B \phi^{(50\Delta ,\tau)-,(\tilde k,\tilde \tau)s}.    
\end{split}
\end{align}
If $\beta^B = 1$, the at-the-money volatility surface is flat. Increasing $\beta^B$ leads to a higher at-the-money convexity. Combining this formula with the previous results leads to the following theorem.
\begin{theorem}\label{thm_no_arbitrage}
\textbf{(No-arbitrage of the volatility surface)} We say that the three-points microscopic volatility surface is arbitrage free if 
\begin{align*}
    \boldsymbol{\phi} \in \mathcal{NA}_3= \{\boldsymbol{\phi}\in \mathcal{M}_{2\times \#\mathcal{K}\times\#\mathcal{T}}(\mathbb{R}), \text{ s.t } \eqref{calendar_spread_sufficient_cond},\eqref{cond_no_arbitrage_butterfly_kernel} \text{ and } \eqref{condition_butterfly_convexity_atm}\text{ are satisfied }  \}.
\end{align*}
\end{theorem}
To summarize:
\begin{itemize}
    \item Condition \eqref{calendar_spread_sufficient_cond} guarantees the absence of calendar spread arbitrage, and reduces the dimension of the Hawkes kernel from $2\times \#\mathcal{K}\times\#\mathcal{T}$ to $2\times \#\mathcal{K}$. It states that, for a fixed strike, the intensity kernel for a different maturity is simply the initial intensity kernel scaled by the square root of the quotient of the two considered maturities. 
    \item Condition \eqref{cond_no_arbitrage_butterfly_kernel} guarantees no-arbitrage for the wings of the volatility surface, ensuring that on average, the behavior of the volatility for a large strike is of $\sqrt{k}$  order of magnitude, where strike $k$ is expressed in log-moneyness. 
    \item Condition \eqref{condition_butterfly_convexity_atm} guarantees convexity of the at-the-money strike which, for slices composed of three strikes, is sufficient to ensure no butterfly arbitrage. 
\end{itemize}
The simplicity of the formulae allows to give simple conditions on the kernel parameters so that the volatility slices are right or left skewed. Consider the following general power-law kernel
\begin{align}\label{power_law_kernel}
    \phi^{(k,\tau)s,(\tilde k,\tilde \tau)\tilde s}(t) = \frac{\alpha^{(k,\tau)s,(\tilde k,\tilde \tau)\tilde s}}{(1+t)^{1+\gamma^{(k,\tau)s,(\tilde k,\tilde \tau)\tilde s}}},
\end{align}
with $(k,\tilde k)\in \{-25\Delta P, 50\Delta, 25\Delta C\}^2,  (\tau,\tilde \tau)\in \mathcal{T}^2, (s,\tilde s)\in \{+,-\}^2$ and assume stationary increments of the Hawkes processes. We have, see \cite{bacry2015hawkes}, for example,
\begin{align*}
    \boldsymbol{\bar \lambda} = \mathbb{E}[\boldsymbol{\lambda}] = \big(I - \boldsymbol{\tilde{\phi}}\big)^{-1}\boldsymbol{\mu}, 
\end{align*}
where $\boldsymbol{\tilde{\phi}}^{(k,\tau)s,(\tilde k,\tilde \tau)\tilde s} = \frac{\alpha^{(k,\tau)s,(\tilde k,\tilde \tau)\tilde s}}{\gamma^{(k,\tau)s,(\tilde k,\tilde \tau)\tilde s}}$. Therefore, on average, $\text{RR}_{25\Delta}>0$ if for every $\tau\in\mathcal{T}, \bar \lambda^{(25\Delta C,\tau)+}-\bar \lambda^{(25\Delta C,\tau)-} > \bar \lambda^{(-25\Delta P,\tau)+}-\bar \lambda^{(-25\Delta P,\tau)-}$ and conversely. This condition can be relaxed in several ways in the spirit of \eqref{cond_no_arbitrage_butterfly_kernel}, thus we obtain the following corollary. 
\begin{corollary}\label{corollary_3points}
\begin{enumerate}\leavevmode
    \item The three-points volatility surface is right (resp. left) skewed if for every $\tau\in\mathcal{T}$
\begin{align*}
    \!\bar \lambda^{(25\Delta C,\tau)+}-\bar \lambda^{(25\Delta C,\tau)-} = \beta^{RR}\big( \bar \lambda^{(-25\Delta P,\tau)+}-\bar \lambda^{(-25\Delta P,\tau)-}\big), \, \beta^{RR}>1 \big(\text{resp. } \beta^{RR}\in(0,1)\big).
\end{align*}
In particular, high $\beta^{RR}$ increases the right skewness of the volatility surface and conversely for $\beta^{RR}$ close to zero.
    \item If $\beta^B$ defined in Equation \eqref{condition_butterfly_convexity_atm} is close to one, the at-the-money volatility surface is flat, while increasing $\beta^B$ leads to a higher at-the-money convexity. 
\end{enumerate} 
\end{corollary}

\subsubsection{Five-points volatility surface}

The volatility surface from the previous section takes into account, for each slice of maturity, the three most liquid traded points denoted by the $-25\Delta P,50\Delta,25\Delta C$ strikes. We can obviously extend this modeling by including for every slice of maturity $\tau$ the volatilities $\sigma^{10\Delta C,\tau},\sigma^{-10\Delta P,\tau}$ corresponding to call and put with strikes such that their delta are equal to $0.10$ and $-0.10$ respectively. We can now define the volatilites of the $10\Delta$ risk reversal and butterfly strategies as
\begin{align*}
        BF_{10\Delta,\tau} = \frac{\sigma^{10D\Delta C,\tau}+\sigma^{-10\Delta P,\tau}}{2}-\sigma^{50\Delta,\tau}, \quad RR_{10\Delta,\tau} = \sigma^{10\Delta C,\tau}- \sigma^{-10\Delta P,\tau}. 
\end{align*}
To ensure convexity of the volatility slice, we must have $BF_{10\Delta,\tau}>0$ which, by analogy with the previous section, leads to the following conditions  
\begin{align}\label{condition_butterfly_convexity_atm_10D}
\begin{split}
    & \frac{\phi^{(10\Delta C,\tau)+,(\tilde k,\tilde \tau)s} + \phi^{(-10\Delta P,\tau)+,(\tilde k,\tilde \tau)s}}{2} = \beta^{\tilde B} \phi^{(50\Delta ,\tau)+,(\tilde k,\tilde \tau)s},\\
    & \frac{\phi^{(10\Delta C,\tau)-,(\tilde k,\tilde \tau)s} +\phi^{(-10\Delta P,\tau)-,(\tilde k,\tilde \tau)s}}{2} = \beta^{\tilde B} \phi^{(50\Delta ,\tau)-,(\tilde k,\tilde \tau)s},    
\end{split}
\end{align}
for $\beta^{\tilde{B}}>1$, which controls the convexity at extreme strikes of the volatility surface. In particular, if $\beta^B <\beta^{\tilde{B}}$, the convexity of the volatility slices increases at extreme strikes. Conversely, if $\beta^{\tilde{B}}$ is close to $\beta^B$, the volatility slice flattens at extreme strikes. This leads to the following theorem, which is the counterpart of Theorem \ref{thm_no_arbitrage} for the five-points volatility surface. 
\begin{theorem}\label{thm_no_arbitrage_five_points}
\textbf{(No-arbitrage of the volatility surface)} We say that the five-points microscopic volatility surface is arbitrage free if 
\begin{align*}
    \boldsymbol{\phi} \in \mathcal{{NA}}_5= \{\boldsymbol{\phi}\in \mathcal{M}_{2\times \#\mathcal{K}\times\#\mathcal{T}}(\mathbb{R}), \text{ s.t } \eqref{calendar_spread_sufficient_cond},\eqref{cond_no_arbitrage_butterfly_kernel},\eqref{condition_butterfly_convexity_atm} \text{ and } \eqref{condition_butterfly_convexity_atm_10D} \text{ are satisfied }  \}.
\end{align*}
\end{theorem}
It is straightforward to find conditions on the coefficients of the kernels $\boldsymbol{\phi}\in \mathcal{{NA}}$ to control the skewness and convexity of the volatility surface. Let us consider the power-law kernel \eqref{power_law_kernel}, for $(k,\tilde k)\in \{-10\Delta P,-25\Delta P, 50\Delta, 25\Delta C,10\Delta C\}^2, (\tau,\tilde \tau)\in \mathcal{T}^2,  (s,\tilde s)\in \{+,-\}^2$ and assume stationary increments of the Hawkes processes, so that we have
\begin{align}\label{expectation_hawkes_power_law_5points}
    \boldsymbol{\bar \lambda} = \mathbb{E}[\boldsymbol{\lambda}] = \big(I - \boldsymbol{\tilde{\phi}}\big)^{-1}\boldsymbol{\mu}, 
\end{align}
where $\boldsymbol{\tilde{\phi}}^{(k,\tau)s,(\tilde k,\tilde \tau)\tilde s} = \frac{\alpha^{(k,\tau)s,(\tilde k,\tilde \tau)\tilde s}}{\gamma^{(k,\tau)s,(\tilde k,\tilde \tau)\tilde s}}$, and \eqref{expectation_hawkes_power_law_5points} is in closed-form. By analogy with Corollary \ref{corollary_3points}, we derive the following results.
\begin{corollary}\label{corollary_5points}\leavevmode
\begin{enumerate}
    \item The five-points volatility surface is right (resp. left) skewed if for every $\tau\in\mathcal{T}$
\begin{align*}
\begin{split}
    & \bar \lambda^{(25\Delta C,\tau)+}-\bar \lambda^{(25\Delta C,\tau)-} \!\!=\! \beta^{RR}_{25\Delta}\big( \bar \lambda^{(-25\Delta P,\tau)+}-\bar \lambda^{(-25\Delta P,\tau)-}\big), \, \beta^{RR}_{25\Delta}>1 \big(\text{resp. } \beta^{RR}_{25\Delta}\in(0,1)\big), \\
    & \bar \lambda^{(10\Delta C,\tau)+}-\bar \lambda^{(10\Delta C,\tau)-} \!\!=\! \beta^{RR}_{10\Delta}\big( \bar \lambda^{(-10\Delta P,\tau)+}-\bar \lambda^{(-10\Delta P,\tau)-}\big), \, \beta^{RR}_{10\Delta}>1 \big(\text{resp. } \beta^{RR}_{10\Delta}\in(0,1)\big).    
\end{split}
\end{align*}
    \item A slice $\tau\in \mathcal{T}$ exhibits a smile (resp. skew) if $RR_{10\Delta,\tau}>RR_{25\Delta,\tau}$ (resp $<$), which is satisfied if
\begin{align*}
    (\beta^{RR}_{10\Delta}-1)\big( \bar \lambda^{(-10\Delta P,\tau)+}-\bar \lambda^{(-10\Delta P,\tau)-}\big) > (\beta^{RR}_{25\Delta}-1)\big( \bar \lambda^{(-25\Delta P,\tau)+}-\bar \lambda^{(-25\Delta P,\tau)-}\big),
\end{align*}
and a sufficient condition is
\begin{align*}
    (\beta^{RR}_{10\Delta}-1)\big( \bar \lambda^{(-10\Delta P,\tau)+}-\bar \lambda^{(-10\Delta P,\tau)-}\big) = \beta^{SS} (\beta^{RR}_{25\Delta}-1)\big( \bar \lambda^{(-25\Delta P,\tau)+}-\bar \lambda^{(-25\Delta P,\tau)-}\big) 
\end{align*}
with $\beta^{SS}\in (1,+\infty)$ (resp. $\beta^{SS}\in (0,1)$). In particular, a high $\beta^{SS}$ leads to a steeper smile and conversely. 
    \item The convexity of the five-points volatility surface increases (resp. decreases) for extreme strikes if, for every $\tau\in\mathcal{T}$, $BF_{10\Delta,\tau}>BF_{25\Delta,\tau}$ (resp.  $BF_{10\Delta,\tau}<BF_{25\Delta,\tau}$). A sufficient condition is 
\begin{align*}
    \beta^{\tilde{B}}> \beta^B (\text{ resp. }\beta^{\tilde{B}}< \beta^B).
\end{align*}
\end{enumerate}
\end{corollary}

\section{Macroscopic limit of the volatility surface}\label{sec_macro_volatility}

We now show the macroscopic behavior of the volatility dynamics. As the absence of calendar spread arbitrage offers a simple parametrization compared to other conditions, in this section, we only assume the absence of calendar spread arbitrage, that is, Condition \eqref{calendar_spread_sufficient_cond} is verified and becomes an equality. Using no butterfly arbitrage would change the Hawkes kernel's coefficients with respect to the strikes, but not the qualitative nature of our results. We adapt the framework used in \cite{tomas2019microscopic}. For every result stated in this section, we refer the reader to \cite{tomas2019microscopic} for corresponding proofs. \medskip

We set $T>0$ and take, in a basis $O$ independent of both $T$ and $t\in[0,T]$, a sequence of triangularizable intensity kernels $\boldsymbol{\phi}^{\mathbf{T}}(t) \in \mathcal{M}_{2\times\#\mathcal{K}\times\#\mathcal{T}}(\mathbb{R})$ with $n_c>0$ non-zero eigenvalues. Using the block matrix notation, the Hawkes kernel can be written as
\begin{align*}
   \boldsymbol{\phi}^{\mathbf{T}}(t) = \mathbf{O} 
   \begin{pmatrix}
    \mathbf{A}^{\mathbf{T}}(t) & 0 \\
    \mathbf{B}^{\mathbf{T}}(t) & \mathbf{C}^{\mathbf{T}}(t)
   \end{pmatrix}
   \mathbf{O}^{-1},
\end{align*}
with $\mathbf{A}^{\mathbf{T}}: \mathbb{R}_+ \rightarrow \mathcal{M}_{n_c}(\mathbb{R})$, $\mathbf{B}^{\mathbf{T}}: \mathbb{R}_+ \rightarrow \mathcal{M}_{2\times\#\mathcal{K}\times\mathcal{T}-n_c,n_c}(\mathbb{R}),\mathbf{C}^{\mathbf{T}}: \mathbb{R}_+ \rightarrow \mathcal{M}_{2\times\#\mathcal{K}\times\mathcal{T}-n_c}(\mathbb{R})$. We impose the following limit condition on the Hawkes kernel:
\begin{align*}
    \mathbf{T}^\alpha \big(\mathbf{I} -\int_0^\infty \mathbf{A}^{\mathbf{T}} \big) \rightarrow_{\mathbf T \rightarrow+\infty} \mathbf{K},
\end{align*}
where $\mathbf{K}$ is an invertible matrix and $\alpha>0$ is the Hurst exponent of the volatility surface. This condition is in fact the saturation of the stability condition of the Hawkes kernel. We finally add the heavy-tail condition on $\mathbf A=\lim_{\mathbf T \rightarrow +\infty} \mathbf{A}^{\mathbf{T}}$, that is
\begin{align*}
    \alpha x^\alpha\int_x^\infty \mathbf A(s) ds \rightarrow_{x\rightarrow +\infty} \mathbf{M},
\end{align*}
with $\mathbf{M}$ being an invertible matrix. \medskip

For $t\in[0,T]$, we define the following rescaled processes:
\begin{align*}
    & \mathbf{X}_t^{\mathbf{T}}=\frac{1}{T^{2\alpha}}\mathbf{N_{tT}^T}, \quad \mathbf{Y}_t^{\mathbf{T}}=\frac{1}{T^{2\alpha}}\int_0^{t\mathbf T} \boldsymbol{\lambda}_{\mathbf{s}} ds, \quad \mathbf{Z}_t^{\mathbf{T}}=T^{\alpha}\big(\mathbf{X_t}^{\mathbf{T}} - \mathbf{Y_t}^{\mathbf{T}} \big) =  \mathbf{M_{tT}^T}, \quad \boldsymbol{\sigma}_t^{\mathbf{T}} = \frac{1}{T^{2\alpha}}\big( \mathbf{N}_{\mathbf{tT}}^{\mathbf {T}+}-\mathbf{N}_{\mathbf{tT}}^{\mathbf {T}-} \big).
\end{align*}
Particularly, $\boldsymbol{\sigma^T}$ is the rescaled microscopic volatility surface. Throughout the section, we assume the following:
\begin{assumption}
For all $(\tilde{k},\tilde{\tau})\in \mathcal{K}\times\mathcal{T}$, and $(s,\tilde{s})\in \{+,-\}^2$, 
\begin{align*}
    \phi^{(k,\tau)s,(\tilde{k},\tilde{\tau})\tilde{s}}(t) = \sqrt{\frac{\tau}{\tilde{\tau}}}\phi^{k,\tilde{k}}(t),
\end{align*}
meaning that the self $(++,--)$ and cross $(+-,-+)$ exciting terms are the same.
\end{assumption}
This assumption can be relaxed and is made only to provide meaningful results for some specific forms of $\boldsymbol{\phi}$. 

\subsection{Limiting processes in the separable, one-dimensional case}

In this section, we consider the case of the separable Hawkes kernel. We assume 
\[
\phi^{k,\tilde{k}}(t) = \tilde z(k) \tilde z(\tilde{k}) \varphi(t) \, ,
\]
where $\varphi \colon \R_+ \to \R_+ $, $\tilde z : \mathcal{K} \to \R_+$  represent the overall intensity for changes of the implied volatility surface depending on time and strike respectively. 
\begin{remark}
In the context of a power-law kernel \eqref{power_law_kernel}, this form of kernel coefficients means that there is a single power coefficient $\gamma^{(k,\tau)s,(\tilde{k},\tilde{\tau})\tilde{s}}=\gamma>0$, and only the scaling coefficients $\alpha^{(k,\tau)s,(\tilde{k},\tilde{\tau})\tilde{s}}$ can differ from one strike to another. 
\end{remark}
Therefore, $\boldsymbol{\phi}$ is a rank one matrix defined by
\[
\bm{\phi} = \sqrt{\tau} \varphi \begin{pmatrix}
\frac{1}{\sqrt{\tau}}\tilde z \tilde z^\top & \frac{1}{\sqrt{\tau}}\tilde z \tilde z^\top & \dots & \frac{1}{\sqrt{\tau}}\tilde z \tilde z^\top\\
\frac{1}{\sqrt{\tau_2}} \tilde z \tilde z^\top &  \frac{1}{\sqrt{\tau_2}} \tilde z \tilde z^\top  & \dots & \frac{1}{\sqrt{\tau_2}} \tilde z \tilde z^\top \\
\vdots \\
\frac{1}{\sqrt{\tau_M}} \tilde z \tilde z^\top & \frac{1}{\sqrt{\tau_M}} \tilde z \tilde z^\top & \dots & \frac{1}{\sqrt{\tau_M}} \tilde z \tilde z^\top
\end{pmatrix} \, .
\]
This matrix has a single non-zero eigenvalue $\normsq{z}^2 (\frac{1}{\sqrt{\tau}}+\frac{1}{\sqrt{\tau_1}} + \cdots + \frac{1}{\sqrt{\tau_M}})$ associated to the eigenvector $v = \frac{1}{\normsq{\tilde z} \sqrt{\frac{1}{\sqrt{\tau}}+\frac{1}{\sqrt{\tau_2}} + \cdots + \frac{1}{\sqrt{\tau_d}}}} (\frac{1}{\sqrt{\tau}}\tilde z , \frac{1}{\sqrt{\tau_2}} \tilde z, \dots, \frac{1}{\sqrt{\tau_M}} \tilde z )$. In order for the Hawkes process to be nearly-unstable we impose
\[
\norm{\varphi} = \frac{1}{\normsq{\tilde z}^2 (1+\frac{\sqrt{\tau}}{\sqrt{\tau_2}} + \cdots + \frac{\sqrt{\tau}}{\tau_M})} \, .
\]
Then, by using the results of \cite{el2018microstructural}, we have that at the limit, the dynamics of the volatility is given by $\boldsymbol{\sigma_t} = F_t v$, where 
\begin{align*}
    F_t &= C \int_0^t \sqrt{V_s} dW_s, \\
    V_t &= \frac{1}{\Gamma(\alpha)} \int_0^t (t-s)^{\alpha - 1} (\theta - V_s) ds + \frac{1}{\Gamma(\alpha)} \int_0^t (t-s)^{\alpha - 1} \lambda \sqrt{V_s} dZ_s,
\end{align*}
$(C,\theta,\lambda)\in \mathbb{R}^3_+$ and $(W,Z)$ is a $2$-dimensional Brownian motion. Note that the value of the coefficients in the macroscopic volatility process depends crucially on the rescaling. In Section \ref{sec_rescaling_general_case}, we provide a characterization of these coefficients in the general case of a triangularizable intensity kernel with the specific rescaling described at the beginning of the section. In the present case, the macroscopic volatility surface is driven by a single factor of risk $F_t$ whose dynamics is that of a rough Heston model. This corresponds to the so-called ``market'' or ``level'' factor described in \cite{cont2002dynamics}. All the components of the eigenvector $v$ are positive. Thus, when factor $F_t$ increases, all the implied volatilites increase and conversely. 

\medskip
Finally, note that the volatility of the macroscopic implied volatility is a rough process. This property does not contradict the studies on the volatility's roughness always treating the realized volatility because, in this article, we model the implied volatility, which has no reason to be rough.

\subsection{Limiting process in the semi-separable, factor case}
In this section, we consider the case of the semi-separable Hawkes kernel. We assume that
\[
\phi^{k,\tilde k}(t) = \sum_{i=1}^{r} \tilde z_i(k) \tilde z_i(\tilde{k})^\top \varphi_i(t) ,
\]
where each factor has an associated kernel in time to reflect lead-lag effects in the overall surface. 
\begin{remark}
In the context of a power-law kernel \eqref{power_law_kernel}, this form of kernel coefficients means that there is a single power coefficient $\gamma_i^{(k,\tau)s,(\tilde{k},\tilde{\tau})\tilde{s}}=\gamma_i>0$ for every of the $F^i_t, i\in \{1,\dots,r\}$ factors and the scaling coefficients $\alpha^{(k,\tau)s,(\tilde{k},\tilde{\tau})\tilde{s}}$ can differ depending on the strike and the considered factor $F^i_t, i\in \{1,\dots,r\}$. 
\end{remark}

We assume that $\tilde z_i, i\in \{1,\dots,r\}$ are orthogonal, \textit{i.e.} $\sum_k \tilde z_i(k) \tilde z_j(\tilde k) = 0$ if $i \neq j$. Therefore, $\bm{\phi}$ is a rank $r$ matrix
\[
\bm{\phi} = \sqrt{\tau}  \sum_{i=1}^{r} \varphi_i \begin{pmatrix}
\frac{1}{\sqrt{\tau}}\tilde z_i \tilde z_i^\top & \frac{1}{\sqrt{\tau}}\tilde z_i \tilde z_i^\top & \dots & \frac{1}{\sqrt{\tau}}\tilde z_i \tilde z_i^\top\\
\frac{1}{\sqrt{\tau_2}} \tilde z \tilde z_i^\top &  \frac{1}{\sqrt{\tau_2}} \tilde z_i \tilde z_i^\top  & \dots & \frac{1}{\sqrt{\tau_2}} \tilde z_i \tilde z_i^\top \\
\vdots \\
\frac{1}{\sqrt{\tau_M}} \tilde z_i \tilde z_i^\top & \frac{1}{\sqrt{\tau_M}} \tilde z_i \tilde z_i^\top & \dots & \frac{1}{\sqrt{\tau_M}} \tilde z_i \tilde z^\top
\end{pmatrix} \, .
\]
Thus, the eigenvectors of the Hawkes kernel are
$v_{i} = \frac{1}{\normsq{\tilde z_i} \sqrt{\frac{1}{\sqrt{\tau}}+\frac{1}{\sqrt{\tau_2}} + \cdots + \frac{1}{\sqrt{\tau_M}}}} (\frac{1}{\tau}\tilde z_i ,\frac{1}{\sqrt{\tau_2}}  \tilde z_i, \dots, \frac{1}{\sqrt{\tau_M}}\tilde z_i )$ with associated eigenvalues $\normsq{z_i}^2 (\frac{1}{\sqrt{\tau}}+\frac{1}{\sqrt{\tau_1}} + \cdots + \frac{1}{\sqrt{\tau_M}})$. They form an orthonormal basis as the $\tilde z_i$ are orthogonal. Therefore, the criticality condition for the eigenvector $v_i$ is:
\[
\norm{\varphi_i} = \frac{1}{\normsq{\tilde z_i}^2 (1+\frac{\sqrt{\tau}}{\sqrt{\tau_2}} + \cdots + \frac{\sqrt{\tau}}{\sqrt{\tau_M}})} \, .
\]
Then, at the limit, we obtain  $\boldsymbol{\sigma_t} = \sum_{i=1}^{r}  v_{i} F^i_t$ where 
\begin{align*}
    F^{i}_t &= C^i \int_0^t \sqrt{V^i_s} dW^i_s, \\
    V^{i}_t &= \frac{1}{\Gamma(\alpha)} \int_0^t (t-s)^{\alpha - 1} (\theta^i - V^i_s) ds + \frac{1}{\Gamma(\alpha)} \int_0^t (t-s)^{\alpha - 1} \lambda^{i} \sqrt{V^i_s} dZ^i_s,
\end{align*}
with $(C^i,\theta^i,\lambda^i)\in \mathbb{R}_+^3$ for all $i\in \{1,\dots,r\}$ and $(W^i,Z^i)_{i\in\{1,\dots,r\}}$ being a $2r$-dimensional Brownian motion. The macroscopic limit of the volatility surface dynamics is given by a sum of risk factors having rough volatility.

\medskip
In this setting, it is easy to parametrize the microscopic kernel to recover the classic ``level-skew-convexity'' behavior of the volatility surface. Assume that $r=3$, that is, there are three factors of risk driving the macroscopic volatility surface.
\begin{itemize}
    \item By setting $\tilde{z}_1(k) > 0$ for all $k\in \mathcal{K}$, the first factor corresponds to the ``level'' factor of the implied volatility surface: when this factor increases, all the implied volatilities increase and conversely. A possible parametrization is then
\begin{align*}
    \tilde{z}_1(k) = \nu k,
\end{align*}
with $\nu>0$, close to zero: the eigenvector components increase linearly with respect to the moneyness, meaning that the ``level'' factor's impact is slightly higher for out-of-the-money calls.   
    \item By setting $\tilde{z}_2(k) >0$ (resp. $\tilde{z}_2(k) <0$) for $k>1$ (resp $k<1$), the second factor corresponds to the ``calendar'' factor of the implied volatility surface: when this factor increases, the implied volatility of out-of-the-money calls increase, while those of out-of-the-money puts decrease. A possible parametrization is then
\begin{align*}
 \tilde{z}_2(k)=c_1\bigg(\frac{1}{1+e^{-c_2(k-1)}}-0.5\bigg),   
\end{align*}    
with $c_1,c_2 >0$. The parameter $c_1$ controls the scale of the factor. The parameter $c_2$ controls the steepness of the factor's change around the money: when $c_2$ increases, $\tilde{z}_2(k)$ changes its sign quicklier around $k=1$ and conversely.    
    \item By setting $\tilde{z}_3(k)$ parabolic with respect to $k$, minimized at $k=1$, the third factor $F^3_t$ corresponds to the ``butterfly'' factor of the implied volatility surface: a variation of this factor leads to a change of convexity of the volatility surface and a downward sloping term structure (which is already incorporated into the intensity kernel through the no calendar spread arbitrage condition \eqref{calendar_spread_sufficient_cond}). A possible parametrization is then
\begin{align*}
    \tilde{z}_3(k)=c_3 (k-1)^2 - c_4, 
\end{align*}
with $c_3,c_4>0$. The parameter $c_3$ scales  the factor and the parameter $c_4$ controls the factor's level. 
\end{itemize}

\subsection{General case}\label{sec_rescaling_general_case}

In this section, we assume no specific form on $\phi^{k,\tilde{k}}(t)= \tilde{Z}(k,\tilde{k},t)$. This covers the case of power-law kernels where 
\begin{align*}
    \phi^{k,\tilde{k}}(t) = \frac{\alpha^{k,\tilde{k}}}{(1+t)^{1+\gamma^{k,\tilde{k}}}},
\end{align*}
with $(\alpha^{k,\tilde{k}},\gamma^{k,\tilde{k}})\in \mathbb{R}^{\star 2}_+$. We can rewrite the kernel in the following basis:
\begin{align*}
       \boldsymbol{\phi}(t) = \mathbf{O} 
   \begin{pmatrix}
    \mathbf{A}(t) & 0 \\
    \mathbf{B}(t) & \mathbf{C}(t)
   \end{pmatrix}
   \mathbf{O}^{-1},
\end{align*}
where 
\begin{align*}
\mathbf{O} &= \begin{pmatrix}
\frac{1}{\sqrt{\tau}}I & \frac{1}{\sqrt{\tau}}I  & \dots & 0 \\
\frac{1}{\sqrt{\tau_2}} I  &  0   & \dots &0  \\
\vdots \\
\frac{1}{\sqrt{\tau_M}}I & -\frac{1}{\sqrt{\tau_M}}I & \dots & -\frac{1}{\sqrt{\tau_M}}I
\end{pmatrix} \boldsymbol{G} = \begin{pmatrix}
 \bf{O_{11}} & \bf{O_{12}} \\
 \bf{O_{21}} & \bf{O_{22}}
\end{pmatrix},
\end{align*}
and $\boldsymbol{G} := \text{diag}(\tilde G, \dots, \tilde G)$, where $\tilde G$ is the basis change which makes $\tilde Z$ tridiagonal for all $t$. Let

\begin{align*}
\allowdisplaybreaks
 &  \bf{O^{-1}} = \begin{pmatrix}
 \bf{O^{-1}_{11}} & \bf{O^{-1}_{12}} \\
 \bf{O^{-1}_{21}} & \bf{O^{-1}_{22}}
\end{pmatrix},  \\
& \bf{\Theta^1} = \big( \bf{O}_{11} + \bf{O}_{12}(\bf{I} - \int_0^\infty \bf{C})^{-1}\int_0^\infty \bf{B}\big)\bf{K}^{-1}, \\
& \bf{\Theta^2} = \big( \bf{O}_{21} + \bf{O}_{22}(\bf{I} - \int_0^\infty \bf{C})^{-1}\int_0^\infty \bf{B}\big)\bf{K}^{-1},  \\
& \boldsymbol{\theta_0} = \begin{pmatrix}
 \bf{O^{-1}_{11}} & \bf{0}, \\
 \bf{0} & \bf{O^{-1}_{22}}
\end{pmatrix} \boldsymbol{\mu}, \\
& \bf{\Lambda} = \frac{\alpha}{\Gamma(1-\alpha)}\bf{KM^{-1}},
\end{align*}
and assume that $\boldsymbol{\phi}$ has $n_c>0$ non-zero eigenvalues. Then, for any limit point $(\bf{X,Y,Z})$ of the sequence $(\bf{X^T,Y^T,Z^T})$, which is $C$-tight for the Skorokhod topology, there exists a positive process $\mathbf{V}$ and a $2\times\#\mathcal{K}\times\#\mathcal{T}$ dimensional Brownian motion $\bf{B}$ such that 
\begin{align*}
    \bf{X_t} = \int_0^t \bf{V_s}ds, \quad \bf{Z_t} = \int_0^t \text{diag}(\sqrt{\bf{V_s}})d\bf{B}_s.
\end{align*}
Moreover there exists a process $\bf{\tilde{V}}$ of Holder regularity $\alpha-\frac{1}{2}-\epsilon$ for any $\epsilon>0$ such that $\bf{\Theta^1\tilde{V}}= (V^1,\dots,V^{n_c}),\bf{\Theta^2\tilde{V}}= (V^{n_c+1},\dots,V^{2\times\#\mathcal{K}\times\#\mathcal{T}})$ and for all $t\in [0,1]$,
\begin{equation*}
    \begin{split}
        \bm{\Tilde{V}}_t &= \dfrac{1}{\Gamma(\alpha)}\bm{\Lambda} \int_0^t (t-s)^{\alpha-1}(\bm{\theta_0} - \bm{\Tilde{V}}_s) ds  \\ &+\dfrac{1}{\Gamma(\alpha)}\bm{\Lambda} \int_0^t (t-s)^{\alpha-1} \bm{O^{(-1)}_{11}}\textnormal{diag}(\sqrt{\bm{\Theta^1} \bm{\Tilde{V}}_s}) d\bm{W^1}_s 
        \\ &+\dfrac{1}{\Gamma(\alpha)}\bm{\Lambda} \int_0^t (t-s)^{\alpha-1} \bm{O^{(-1)}_{12}}\textnormal{diag}(\sqrt{\bm{\Theta^2} \bm{\Tilde{V}}_s}) d\bm{W^2}_s,
    \end{split}
\end{equation*}
where $\bm{W^1} := (B^1, \cdots, B^{n_c})$, $\bm{W^2} := (B^{n_c+1}, \cdots, B^{2\times\#\mathcal{K}\times\mathcal{T}})$, $\bm{\Theta^1}$, $\bm{\Theta^2}$, $\bm{O^{(-1)}_{11}}$, $\bm{O^{(-1)}_{12}}, \bm{\theta_0}$ do not depend on the chosen basis.
Finally, any limit point $\boldsymbol{\sigma}$ of the rescaled volatility surface $\boldsymbol{\sigma}^T$ satisfies
\begin{align}\label{vol_rescaled_general_case}
\boldsymbol{\sigma_t} = (\bm{I} + \bm{\Delta}) \bm{Q}^\top (\int^t_0 \textnormal{diag}(\sqrt{\bm{V_s}}) d\bm{B_s} + \int_0^t \bm{\mu_s} ds),    
\end{align}
where 
\begin{align*}
& \Delta_{(k,\tau),(\tilde{k},\tilde{\tau})} = \underset{T \to \infty}{\lim}\Big( \norm{\psi^{T}_{(\tilde{k},\tilde{\tau})+,(k,\tau)+}} - \norm{\psi^{T}_{(\tilde{k},\tilde{\tau})-,(k,\tau)+}} \Big),   \\
& \mathbf{Q}=\big(e_1-e_2|\dots|e_{2\times\#\mathcal{K}\#\mathcal{T}-1}-e_{2\times\#\mathcal{K}\#\mathcal{T}}),
\end{align*}
and $(e_i)_{1\leq i\leq 2\times\#\mathcal{K}\#\mathcal{T}}$ is the canonical basis of $\mathbb{R}^{2\times\#\mathcal{K}\#\mathcal{T}}$. 

\medskip
We refer to \cite{tomas2019microscopic} for the proof of this theorem. In this setting, the macroscopic volatility surface is driven by $2\times\mathcal{K}\times\mathcal{T}$ sources of risk having rough volatilities. The first $n_c$ sources of risk $\mathbf{W^1}$ come from the non-zero eigenvalues of the Hawkes kernel $\boldsymbol{\phi}$ whereas the $2\times\mathcal{K}\times\mathcal{T}-n_c$ others come from the zero-valued eigenvalues. This result contrasts with previous sections where the intensity kernel was diagonalizable, and the only sources of risk corresponded to the non-zero eigenvalues. 

\section{Applications}\label{sec_applications}

\subsection{Backtesting of option market-making strategies}

\subsubsection{Description}
Option market-making is known to be the new ``hot'' topic in the industry. Indeed, it accounts for a considerable part of the trading revenue for proprietary shops and hedge funds. In the academic literature, option market-making models have been studied in \cite{baldacci2019algorithmic, baldacci2020approximate, el2015stochastic, stoikov2009option}. In \cite{el2015stochastic,stoikov2009option}, the authors consider a single-option market driven by a stochastic volatility model and assume that the position is always $\Delta$-hedged. They provide optimal bid and ask quotes for the options and focus either on the risk of model misspecification or the residual risk due to the presence of stochastic volatility. In \cite{baldacci2019algorithmic,baldacci2020approximate}, the authors propose several models to deal with the market-making of a large portfolio of options. The authors make a constant Greek assumption in the former, which is well suited for market-making on long-dated options: the only variable of interest is the aggregated portfolio, where each position is weighted by its Vega. In the latter, the authors make an approximation directly on the form of the value function of the market-maker's optimization problem. Thus, the Greeks are allowed to vary with time, spot, and stochastic volatility, and one can manage each position individually, even for a very large portfolio. \medskip

These models enable to design market-making strategies on options. However, they all make the assumption that the number of buy (resp. sell) filled limit orders on option $(k,\tau)$ on $[0,t]$ is a point process $N^{(k,\tau)+}$ (resp. $N^{(k,\tau)-}$) without self- or cross-exciting properties. In other words, the intensity of $N^{(k,\tau)+}$ is only a function of the time $t$, the spot $S$, the stochastic volatility $\nu$ and the spread $\delta^{(k,\tau)+}$ quoted by the market-maker on the ask side of option $(k,\tau)$. If we assume $\mathbb{P}=\mathbb{Q}$, that is the risk-neutral measure is calibrated only using historical data, the intensities are functions only of time and spread, and are denoted by \[\Lambda^{(k,\tau)s}_t(\delta_t^{(k,\tau)s}), \quad (k,\tau)\in \mathcal{K}\times\mathcal{T}, s\in \{+,-\}.\]
In this case, it is particularly easy to backtest the market-making strategy obtained in \cite{baldacci2019algorithmic,baldacci2020approximate}. We denote the inventory on option $(k,\tau)$ by $q^{(k,\tau)}$ and the optimal strategy of the market-maker on option $(k,\tau)$ by 
\begin{align}\label{optimal_trading_MM}
\delta^{\star (k,\tau)+}(t,q_t),\delta^{\star (k,\tau)-}(t,q_t),
\end{align}
where $q_t=(q_t^{(k,\tau)})_{(k,\tau)\in\mathcal{K}\times\mathcal{T}}$.\medskip

In order to backtest this optimal strategy, we simulate the number of filled buy (resp. sell) limit orders on option $(k,\tau)$ using a point process $\bar{N}^{(k,\tau)+}$ (resp. $\bar{N}^{(k,\tau)-}$) with intensity 
\begin{align}\label{intensity_hawkes_backtest}
    \bar{\lambda}^{(k,\tau)s}\big(t,\delta^{\star (k,\tau)s}(t,q_t)\big) = \lambda_t^{(k,\tau)s}\Lambda^{(k,\tau)s}_t(\delta^{\star (k,\tau)s}(t,q_t)),
\end{align}
where 
\begin{align*}
    \lambda_t^{(k,\tau)s} = \mu^{(k,\tau)s} + \sum_{(\tilde k,\tilde \tau)}\sum_{\tilde{s}\in \{+,-\}}\int_0^t \phi^{(k,\tau)s,(\tilde k,\tilde \tau)\tilde{s}}(t-u)dN^{(\tilde k,\tilde \tau)\tilde{s}}_u,
\end{align*}
and $q_t^{(k,\tau)} = \bar{N}_t^{(k,\tau)+} - \bar{N}_t^{(k,\tau)-}$. Thanks to Equation \eqref{intensity_hawkes_backtest}, we incorporate the strategy of the market-maker in the intensities of the Hawkes processes which represent the market flow on each option. Thus, we can backtest the trading strategy in a more realistic framework. 

\subsubsection{Numerical results}

We conduct a backtest of the optimal market-making strategy in the context of the three point volatility surface described previously. We choose a single slice of maturity $\mathcal{T}=\{\tau\}=1$ week and three options of strike (expressed in log-moneyness)
\begin{align*}
    k_1 = 0.98,\quad  k_2 = 1,\quad  k_3 = 1.02. 
\end{align*}
The influence of the market-maker's quotes on the intensity of orders' arrival are given by 
\begin{align*}
    \Lambda^{(k,\tau)+}_t(\delta) = \Lambda^{(k,\tau)-}_t (\delta) = \frac{\lambda^{k,\tau}}{1+\exp\big(\alpha + \frac{\beta}{\mathcal{V}^{(k,\tau)}}\delta\big)},
\end{align*}
where $\lambda^{k,\tau}=\frac{252\times 70}{1+ 8\times |k-1|}, \alpha = -0.7, \beta = 10$. The choice of $\lambda^{k,\tau}$ corresponds to $70$ requests per day for the at-the-money option, and decreases to $27$ for far-from-the-money options. The choice of $\alpha$ corresponds to a probability of $\frac{1}{1+e^{-0.8}}\approx 66\%$ to trade when the answered quote is the mid-price $(\delta=0)$. The choice of $\beta$ corresponds to a probability of $\frac{1}{1+e^{-0.8}}\approx 68\%$ to trade when the answered quote corresponds to an implied volatility $1\%$ better for the client and a probability of $\frac{1}{1+e^{-0.6}}\approx 64\%$ to trade when the answered quote corresponds to an implied volatility $1\%$ worse for the client. We assume transactions of unitary volume, a trading horizon of $T=1$ day and a risk-aversion parameter of $2\times 10^{-5}$ for the market-maker.\footnote{The extension to multiple transaction sizes is straightforward using marked point processes, see for example \cite{baldacci2019algorithmic}.} Finally, we set $\mu^{(k,\tau)s}= 1$ for all $(k,\tau)\in\mathcal{K}\times\mathcal{T},s\in\{+,-\}$ and the Hawkes kernel is 
\begin{align*}
    & \phi^{(k_1,\tau)s,(k_2,\tau)\tilde{s}} = \phi^{(k_2,\tau)s,(k_1,\tau)\tilde{s}} = \phi^{\text{ITM-ATM}}, \\
    & \phi^{(k_1,\tau)s,(k_3,\tau)\tilde{s}} = \phi^{(k_3,\tau)s,(k_1,\tau)\tilde{s}} = \phi^{\text{OTM-ATM}}, \\
    & \phi^{(k_2,\tau)s,(k_3,\tau)\tilde{s}} = \phi^{(k_3,\tau)s,(k_2,\tau)\tilde{s}} = 0, \\
    & \phi^{(k_1,\tau)s,(k_1,\tau)\tilde{s}} = \phi^{\text{ITM}}, \\
    & \phi^{(k_2,\tau)s,(k_2,\tau)\tilde{s}} = \phi^{\text{ATM}}, \\
    & \phi^{(k_3,\tau)s,(k_3,\tau)\tilde{s}} = \phi^{\text{OTM}}, \vspace{-10mm}
\end{align*}
for all $(s,\tilde{s})\in \{+,-\}$, and for all $t\in[0,T]$,
\begin{align*}
    &  \phi^{\text{ITM-ATM}}(t) = \frac{\alpha^{\text{ITM-ATM}}}{(1+t)^{1+\gamma^{\text{ITM-ATM}}}}, \quad \alpha^{\text{ITM-ATM}} = 0.18, \gamma^{\text{ITM-ATM}} = 0.15,  \\
    &  \phi^{\text{OTM-ATM}}(t) = \frac{\alpha^{\text{OTM-ATM}}}{(1+t)^{1+\gamma^{\text{OTM-ATM}}}}, \quad \alpha^{\text{OTM-ATM}} = 0.13, \gamma^{\text{OTM-ATM}} = 0.15, \\
    & \phi^{\text{ITM}}(t) = \frac{\alpha^{\text{ITM}}}{(1+t)^{1+\gamma^{\text{ITM}}}}, \quad \alpha^{\text{ITM}} = 0.48 , \gamma^{\text{ITM}} =0.08, \\
     & \phi^{\text{ATM}}(t) = \frac{\alpha^{\text{ATM}}}{(1+t)^{1+\gamma^{\text{ATM}}}}, \quad \alpha^{\text{ATM}} = 0.52, \gamma^{\text{ATM}} =0.08,  \\
    & \phi^{\text{OTM}}(t) = \frac{\alpha^{\text{OTM}}}{(1+t)^{1+\gamma^{\text{OTM}}}}, \quad \alpha^{\text{OTM}} = 0.14, \gamma^{\text{OTM}} =0.08.
\end{align*}
We assume power law intensity kernels and no cross-excitation between in-the-money and out-of-the-money options. For sake of simplicity, we assume symmetry between buy and sell self and cross intensities. Moreover
\begin{itemize}
    \item $\phi^{\text{ITM-ATM}}>\phi^{\text{OTM-ATM}}$ : there is more cross excitation between the in-the-money and at-the-money options than between the out-of-the-money and at-the-money options.
    \item $\phi^{\text{ATM}} > \phi^{\text{ITM}} > \phi^{\text{OTM}}$: the self excitation of buy or sell orders is the highest on at-the money options and the lowest on out-of-the-money options. 
\end{itemize}
\vspace{-3mm}
\begin{figure}[H]
\centering
\includegraphics[width=9.5cm]{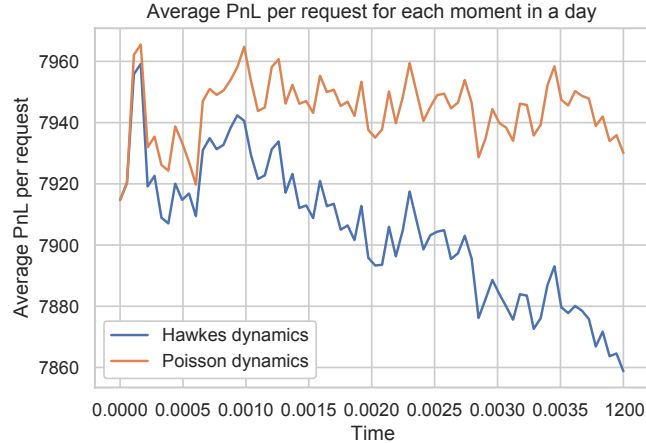}
\vspace{-2mm}
\caption{PnL with respect to time (in seconds) under the Poisson and Hawkes assumption.}
\label{fig::pnl}
\end{figure}
We show in Figure \ref{fig::pnl} the Profit and Loss (PnL) obtained using the optimal strategy \eqref{optimal_trading_MM} when we perform a backtest with the assumption that the intensity of orders' arrival is driven by Hawkes processes compared to the one obtained with Poisson processes assumption (corresponding to the case $\boldsymbol{\phi}=0$). 

\subsection{Market impact curves}

\subsubsection{Description}

Another important application of our framework is the derivation of market impact curves for a given trading strategy at the microscopic scale. We take again the example of option market-making with optimal trading strategy given by \eqref{optimal_trading_MM}. We use the rescaling results of Section \eqref{sec_rescaling_general_case} on the volatility surface:
\begin{align*}
(\bar{\sigma}_t^{(k,\tau)})_{t\in [0,T]} = (\bar{N}_t^{(k,\tau)+}-\bar{N}_t^{(k,\tau)-})_{t\in [0,T]}, 
\end{align*}
where $\bar{N}_t^{(k,\tau)+},\bar{N}_t^{(k,\tau)-}$ have intensities defined by \eqref{intensity_hawkes_backtest}, to obtain an impacted volatility surface $(\boldsymbol{\bar{\sigma}_t})_{t\in[0,T]}$ in the form of \eqref{vol_rescaled_general_case}. Thus, we can compute the market impact of the market-making strategy \eqref{optimal_trading_MM} on the volatility surface at time $t\in[0,T]$ as 
\begin{align}\label{market_impact_whole_curve}
    \text{MI}(t) = \mathbb{E}\big[\|\boldsymbol{\bar{\sigma}_t} - \boldsymbol{\sigma_t}\|_1\big],
\end{align}
where $\boldsymbol{\sigma_t}$ corresponds to the rescaled volatility surface in absence of trading activity, meaning that the Hawkes processes driving the corresponding microscopic volatility surface have intensity 
\begin{align*}
    \lambda_t^{(k,\tau)s} = \mu^{(k,\tau)s} + \sum_{(\tilde k,\tilde \tau)}\sum_{\tilde{s}\in \{+,-\}}\int_0^t \phi^{(k,\tau)s,(\tilde k,\tilde \tau)\tilde{s}}(t-u)dN^{(\tilde k,\tilde \tau)\tilde{s}}_u. 
\end{align*}
For a given trading strategy, we can therefore compute the cumulated impact on the whole volatility surface using \eqref{market_impact_whole_curve}. We can also compute the market impact of the trading strategy on a specific point $(k,\tau)$ of the surface, that is:
\begin{align*}
    \text{MI}^{(k,\tau)}(t) = \mathbb{E}\big[\boldsymbol{\bar{\sigma}^{(k,\tau)}_t} - \boldsymbol{\sigma^{(k,\tau)}_t}\big]. 
\end{align*}
Thus, depending on the trading strategy, we are able to compare its influence on specific parts of the volatility surface. It is of particular interest for a desk of systematic option trading wishing to estimate the PnL of the strategies. 

\subsubsection{Numerical results}

In the classical optimal trading or optimal market-making models, the price processes of stock or options are assumed to be independent from the trading activity. One extension, see \cite{gueant2013dealing}, consists in adding a linear temporary impact, that is the price of an option $(k,\tau)$ is given by
\begin{align*}
    d\tilde{C}^{k,\tau}_t = dC^{k,\tau} + \xi^{k,\tau}\big( d\bar N_t^{(k,\tau)+} - d\bar N_t^{(k,\tau)-}\big), \quad \xi^{k,\tau}>0,
\end{align*}
where the dynamics of $C^{k,\tau}$ is the one of a stochastic volatility model, see \cite{baldacci2019algorithmic}. In the case of Poisson process, that is $\lambda_t^{(k,\tau)s}=0$ for all $(k,\tau)\in \mathcal{K}\times\mathcal{T}, s\in \{+,-\}$, there is no cross-impact: for example, buy orders on option $(k_1,\tau)$ do not change the price of option $(k_2,\tau)$. We use the same model parameters as in the previous section, and set $\xi^{k,\tau}=\frac{5\cdot 10^{-4}}{{C}^{k,\tau}_0}$ for all $k\in \mathcal{K}$, meaning that a buy (resp. sell) order increases (resp. decreases) the price of the option by five basis points. We compute the cross-impact curves of option $k_1$ on the prices of options $k_2$ and $k_3$. \medskip

\begin{figure}[H]
\centering
\includegraphics[width=9.5cm]{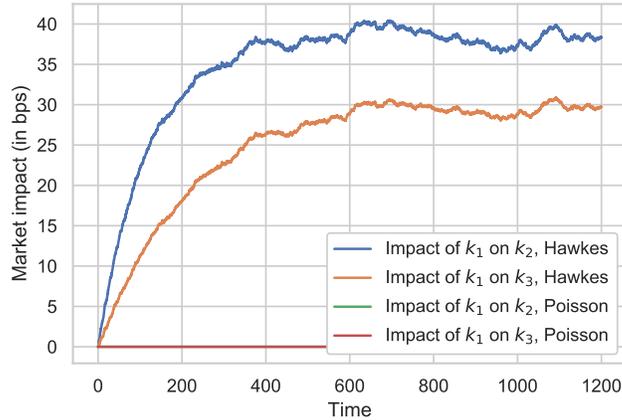}
\vspace{-3mm}
\caption{Cross impact curves (in basis points) with respect to time (in seconds) in the Poisson and Hawkes cases.}
\label{fig::crossimpact}
\end{figure}

We show in Figure \ref{fig::crossimpact} the results in the case of Poisson and Hawkes intensities. The Poisson processes are unable to reproduce the important stylized fact of options markets that buying or selling an option impacts the whole volatility surface. As $\phi^{\text{OTM-ATM}} < \phi^{\text{ITM-ATM}}$, the impact on option $k_2$ of trading option $k_1$ is obviously higher than on option $k_3$. 
\vspace{-3mm}

\section{Conclusion}

We designed a Hawkes-based model for the dynamics of the implied volatility at the microscopic scale. The Hawkes kernel coefficients control the skew and convexity of the volatility surface, and we present sufficient conditions to ensure the absence of arbitrage. These conditions give simple parametrization of the kernel and therefore reduce the number of coefficients to estimate. At the scaling limit, we use some existing results to show that the macroscopic volatility surface dynamics are a sum of risk factors having rough volatility. We show how to parametrize the Hawkes kernel to recover the classic ``level-skew-convexity'' behavior of the volatility surface at the macroscopic limit. Finally, we conduct a backtest of systematic option market-making strategies using our framework and compute the associated market impact on the volatility surface.\medskip

This work opens doors for several extensions: introduction of interactions between the moves of the spot and the volatility surface and its macroscopic extension, study of the scaling limit of the volatility surface when accounting for different Hurst exponents for different options' moneyness. These extensions would lead to new no-arbitrage conditions on the volatility surface.

\bibliographystyle{abbrv}
\bibliography{biblio.bib}

\end{document}